\documentclass{spie}
          \usepackage{graphics}
          
\begin{document}

           \title{The Theory Experiment Connection:
           Rn Space And Inflationary Cosmology}
          \author{Paul Benioff \\
           Physics Division, Argonne National Lab, \\
           Argonne, IL 60439; \\ e-mail: pbenioff@anl.gov}

            \maketitle

            \begin{abstract}
            Based on a discussion of the theory experiment
            connection, it is proposed to tighten the connection
            by replacing the real and complex number basis of
            physical theories by sets $R_{n},C_{n}$ of length
            $2n$ finite binary string numbers.  The form of the numbers
            in $R_{n}$ is based on the infinite hierarchy of $2n$ figure
            outputs from  measurements of any physical quantity
            with an infinite range (distance, energy, etc.).   A
            space and time based on these numbers is described.
            It corresponds to an infinite sequence of
            spherical scale sections $R_{n,\underline{e}}$
            ($\underline{e}$ an integer). Each section has the same
            number of points but the size increases exponentially
            with increasing $\underline{e}.$ The sections converge towards
            an origin which is a space singularity. Iteration of a
            basic order preserving transformation, $F_{<}$ or its inverse
            shows exponential expansion or contraction of the space with
            the origin as a source or sink of space points.
            The suitability of $R_{n}$ space as a framework for
            inflationary cosmology is based on a constant iteration
            rate for $F_{<}$ and making $\underline{e}$ and
            $n=n(t)$ time dependent.  At $t=0$ all space is
            restricted to a region of scale sections
            $R_{n_{0},\underline{e}}$ with $n_{0}$
            (small) and $\underline{e}\leq \underline{e}_{0}$ (negative).
            Inflation, which occurs naturally, is
            stopped automatically at time $t_{I}$ by increasing
            $n$ from $n_{0}$ to $n_{I}>>n_{0}.$ Here $n_{I}$ is
            required to be large enough so that the outermost
            $\Delta$ scale sections of $R_{n_{0}}$ space, which are expanding
            away from the origin at velocities $>c$ at time $t_{I}$,  are
            contained in the scale section $R_{n_{I},0}$ of
            $R_{n_{I}}$ space. This is needed if $R_{n_{I},0}$
            space is to be similar to the usual $R$ space.
            Hubble expansion and the redshift
            are accounted for by a continuing slow increase in
            $n$.  Comparison with experimental data suggests that
            the rate of increase must be at least one $n$ unit every
            $30-60$ million years.
            \end{abstract}
            \keywords{Measurement hierarchy, String numbers, Scale
            invariant space time, Inflationary cosmology}

            \section{Introduction}One important foundational issue
            for physics is the relationship between mathematics
            and physics.  If mathematical objects have
            an ideal objective existence, outside and independent of
            space time, and physical systems exist in and determine
            the properties of space time, then why is mathematics
            relevant to physics? This is not a new problem as shown
            by the title of a paper by
            Wigner \cite{Wigner}: "On the unreasonable
            effectiveness of mathematics in the physical
            sciences."

            This and related questions emphasize the need to
            develop a coherent theory that treats physics and
            mathematics together and not as separate types of
            entities. An important aspect of relating physical
            and mathematical entities in such a theory is to
            recognize the physical nature
            of language \cite{BenLP}.  That is, physical
            representations of language expressions must exist.
            Without them communication, thinking would not be
            possible. Examples include written text, speech
            optical signals, etc..

            A basic aspect of language is that all language expressions
            are represented as symbol strings over some alphabet.
            Physically, expressions
            must be representable by states of one dimensional
            physical systems. As numbers are also finite strings of
            digits in some basis, they also must be representable
            by states of one dimensional physical systems.

            There are many examples of these representations.
            Representations of numbers as product states of qubits
            and their physical realization in quantum computers
            is one. Another is the representation of numbers in
            regular computers by sequences of regions of different
            magnetization or capacitance.  These and other
            representations all illustrate the stringy nature of
            physical representations of language expressions and
            numbers.

            Another aspect that is relevant to developing a
            coherent theory of physics and mathematics is the
            disconnect between theory and experiment. To
            understand this one notes that theoretical predictions
            are given as equations whose solutions are real
            numbers. As real numbers can also be
            represented by infinite digit strings, one sees that
            most real numbers have no names. It follows that the
            real numbers predicted by theory are  nameable real numbers.
           As finite language expressions the  equations that are theoretical
           predictions are names for real numbers.

           This is to be contrasted with the form of numerical
           outcomes of experiments.  These correspond to  length $n$ digit
           string numbers where typically $n$ is small $\sim
           2-10.$  This contrast is the source of the theory
           experiment disconnect.  Predictions are equation names
           of real numbers. Experimental outcomes are short
           finite string digit representations. Its a long way
           from short digit strings to names for real numbers or
           infinite digit strings.

           Computers play an essential role in bridging the
           disconnect gap.  In essence they function as
           translators that translate equation names of real
           numbers to length $m$ digit strings. These strings are supposed
           to be the first $m$ digits of the number named by the
           equation. It is much easier to compare an $n$ digit experimental
           outcome with an $m$ digit prediction than with an equation name.

           One possible method of tightening the theory
           experiment connection is based on replacing the real and complex
           numbers, $R$ and $C$, which are the basis of physical theories, by
           length $n$ string real numbers $R_{n}$ and complex numbers
           $C_{n}=R_{n},I_{n}$ where $I_{n}=iR_{n}$. As a result physical
           theories become mathematical structures over $R_{n}$ and $C_{n}.$

           Consequences of this replacement include the need to adapt the
           theories to the discreteness of $R_{n}$ and $C_{n}$. Also
           a cosmological time dependence of the parameter $n$, which will be used,
           introduces a time dependence into physical theories. However the time
           dependence of $n=n(t)$ will be such that the time dependence of
           physical theories is completely unobservable and is important
           only during the early phases of evolution of the universe.

           In this paper consequences of this replacement will be
           limited to examination of some properties of space and time based on
           these numbers, $R_{n}$ space and time.  Included are the
           observations that $R_{n}$ space and time are scale
           invariant and contain many singular points. It will be
           seen that, by introducing a time dependence of $n=n(t)$
           and of the scale factors,  $R_{n}$ space seems to be
           a useful framework to describe inflationary cosmology
           and the Hubble expansion red shift. Also the time
           dependence of $R_{n}$ space will be such that at the
           present time and during  much of the past, $R_{n}$ space is
           experimentally indistinguishable from the usual $R$
           space.

           It must be emphasized that the choice of the dynamics
           made here is simple and arbitrary. It is made just to show
           the suitability of $R_{n}$ space and time as a
           framework to describe inflationary cosmology and the
           red shift.  It is not based on or derived from any
           physical theory.

           \section{The Numbers $R_{n}$}
           \label{TNR}

           Many choices are possible for the form of the numbers
           in $R_{n}$. Included are the usual string forms used in
           computations or to give the values of physical
           parameters as $\underline{s}_{1}.\underline{s}_{[2,n]}
           \times 10^{w}.$ Here $\underline{s}$ is a length $n$ string
           of decimal digits with the decimal point between the first and
           second digit and $\underline{s}_{1}\neq 0,$ and $w$ is
           an integer.

           The form of the numbers chosen here will be quite
           different as it is based on the outcomes of
           measurements of continuous physical variables with
           infinite ranges, such as distance, time, energy,
           momentum, etc.. Outcomes of $n$ figures in the binary basis
           for these measurements typically have the form,
           $\underline{0}_{[1,n]}$ (nondetect),
           $\underline{0}_{[1,n-1]}1_{n}$ (detect), and
           $\underline{1}_{[1,n]}$ (maximum measurable).

\begin{figure}[t]\begin{center}
           \resizebox{120pt}{120pt}{\includegraphics[280pt,200pt]
           [480pt,400pt]{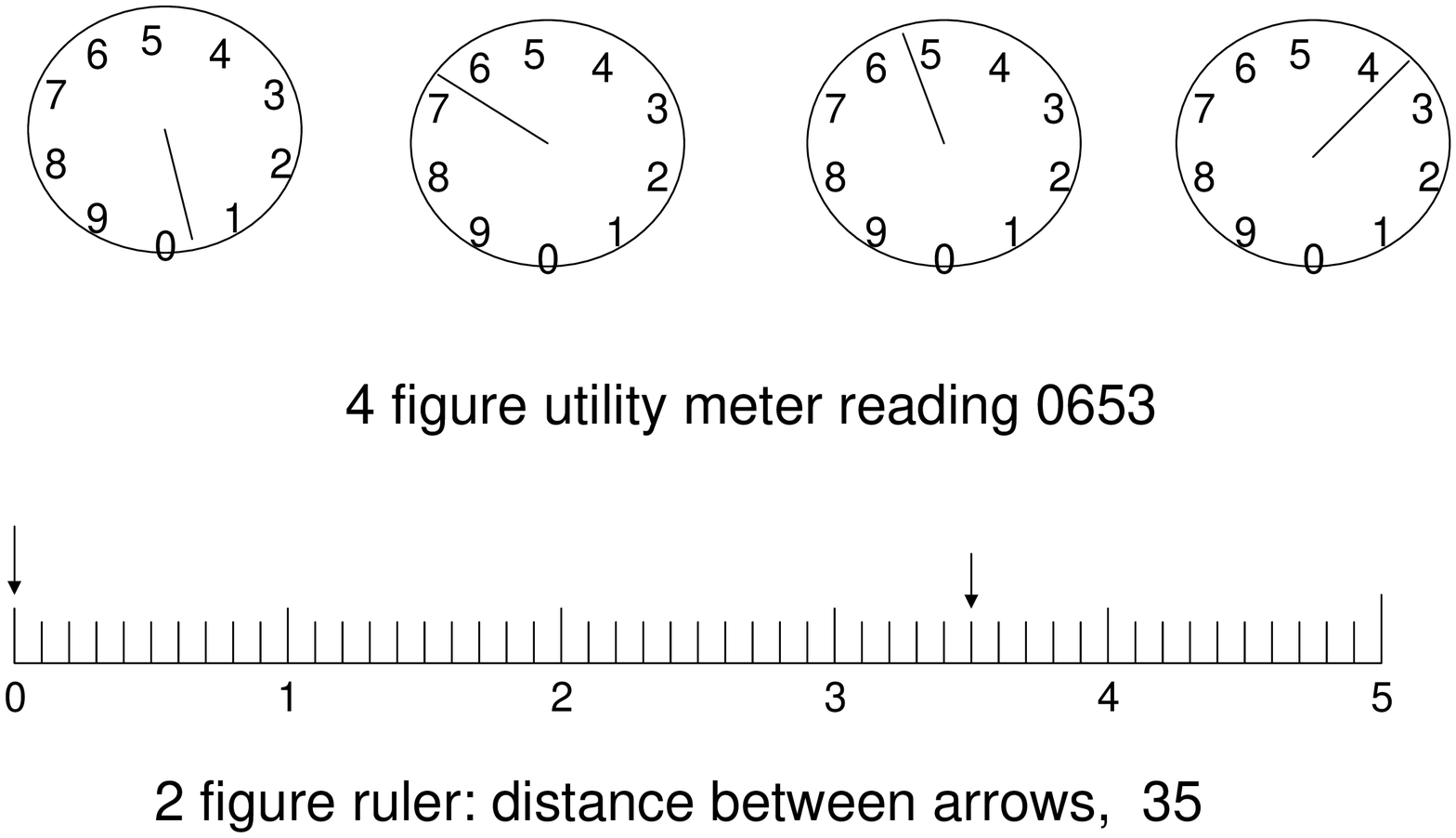}}\end{center}\vspace{0.6in}
           \caption{Utility meter and ruler components of hierarchies
           of measurement apparatuses for measuring physical quantities with
           an infinite range. The position of the lines in each of the
           four meter dials is an example of  the actual output of the
           meter. This is read as a four digit output as shown.}\label{1}
           \end{figure}

           Examples of this type of measurement include
           gas and electric utility meters for measuring
           current and rulers for measuring distance.  These are
           shown in Figure \ref{1}.  Note that here both leading
           and trailing $0s$ are significant.

           Extension of this type of measurement to cover an
           infinite range needs a hierarchy of
           measurement apparatuses that is infinite in both
           directions. An example of the hierarchy for distance measurements
           with $n=2$ would be apparatuses measuring $1$ meter
           in cm units, measuring $100$ meters in $1$ meter units,
           measuring $1$ cm in $.01$ cm units, etc..

           The general relations between the measurements in
           the hierarchy are shown below where the $j+1st$ begins
           just above the range of the $jth$:
           $$\begin{array}{ccccc}\infty\cdots & j+1 & j & j-1& \cdots
           -\infty \\\mbox{} & \left\{\begin{array}{c}\mbox{nondetect}
           \\ \underline{0}_{[1,n]}\end{array}\right \} & \underline{1}_{[1,n]}
           & \mbox{offscale}&\mbox{} \\\mbox{}&\underline{0}_{[1,n-1]}1_{n} &\left
           \{\begin{array}{c}\mbox{off scale} \\
           \underline{1}_{[1,n]}+\underline{0}_{[1,n-1]}1_{n}
           \end{array}\right \} & \mbox{offscale}&\mbox{} \\
           \mbox{}&\mbox{nondetect} & \mbox{nondetect} &
           \underline{1}_{[1,n]}&\mbox{}.\end{array}$$

           Such a hierarchy of measurements may seem
           counterintuitive at first.  However it is supported by
           properties of actual measurements. In particular the
           number of significant figures $n$ in measured values is almost
           independent of the magnitude of the measured quantity.
           For example the number of significant figures in
           distance measurements is limited to a range $2\leq
           n\leq\sim 10$ for magnitudes ranging from Fermis ($10^{-13}\; cm$) to
           billions of light years, a range of $\sim 10^{40}.$

           Based on these considerations the numbers in $R_{n}$
           are taken to have the symmetric form\footnote{From now
           on the sequence length is given  as $2n.$  The label
           $n$ on $R_{n}$ is kept.},
           \begin{equation}\label{num}\begin{array}{c}
           (\pm\underline{s}_{[1,n]}.\underline{s}_{[n+1,2n]},
           2n\underline{e})\; =\pm\underline{s}_{[1,n]}.\underline{s}_{[n+1,2n]}\times
           2^{2n\underline{e}}\\\underline{s}_{[1,2n]}
           \neq\underline{0}_{[1,2n]}\mbox{ if }\underline{e}\neq
           0.\end{array}\end{equation}Here
           $\underline{s}_{[1,n]}.\underline{s}_{[n+1,2n]}$ is a length $2n$
           binary string and $\underline{e}$ is any integer.
           An equivalent representation in a more familiar form is given
           by\begin{equation}\label{numequ}
           (\pm\underline{s}_{[1,n]}.\underline{s}_{[n+1,2n]},2n\underline{e})
           =(\pm\sum_{j=1}^{2n}\underline{s}_{[1,2n]}(j)\times2^{n-j})
           \times2^{2n\underline{e}}\end{equation}.

           Decimal number examples for $n=1$ and $n=2$ with
           $\underline{e}=-1,0,1$ are shown below. $$\begin{array}{ccccc}\mbox{} &
           \mbox{}& \mbox{$n=1$} & \mbox{}& \mbox{} \\
           \cdots & \underline{e}=-1 & \underline{e}=0 & \underline{e}=+1 &
           \cdots \\\cdots & 1/8,\; 2/8,\; 3/8 & 1/2,\; 1,\; 3/2 & 2,\; 4,\; 6 &
           \cdots \\ \mbox{} & \mbox{}& \mbox{$n=2$} & \mbox{}& \mbox{} \\
           \cdots & \underline{e}=-1 & \underline{e}=0 & \underline{e}=+1 &
           \cdots\\ \cdots & 1/64,\; 2/64,\cdots ,15/64 & 1/4,\; 2/4,\cdots
           15/4 & 4,\; 8,\cdots 60 & \cdots \end{array}$$

           The ordering of these numbers can be expressed in
           general as an infinite alternating sequence of
           $2^{2n}-1$ numbers with constant spacing
           $2^{n(2\underline{e}-1)}$ separated by exponential jumps
           of $2^{2n}$ with $\underline{e}\rightarrow
           \underline{e}\pm 1.$  There is no least or greatest
           number.

           Arithmetic with these numbers is somewhat similar to
           computer arithmetic in that roundoff is used.
           This is especially the case  for small $n$ and for
           addition of numbers with different scale factors, as in $(n=2)$
           $$\begin{array}{l}00.10\times 2^{8}+00.10\times
           2^{4}=00.10\times 2^{8}\\10.10\times 2^{4}+11.10\times
           2^{4}=00.01\times 2^{8}.\end{array}$$  Roundoff is
           important also in multiplication as in
           $$\begin{array}{l}(01.01\times 2^{4})\times (11.01\times
           2^{8})=00.01\times 2^{16}\\(00.11\times 2^{4})\times (10.01\times
           2^{-8})=01.11\times 2^{-4}.\end{array}.$$

           \section{$R_{n}$ Space and Time}\label{RST}

           $R_{n}$ space and time is a space time whose locations
           and distances between locations
           are based on the numbers in $R_{n}.$  As such its properties are
           quite different from the usual continuum based space
           time as it inherits many of the properties of $R_{n}.$
           It is discrete and it is scale invariant.

\begin{figure}[h]\begin{center}
           \resizebox{100pt}{80pt}{\includegraphics[330pt,270pt]
           [500pt,400pt]{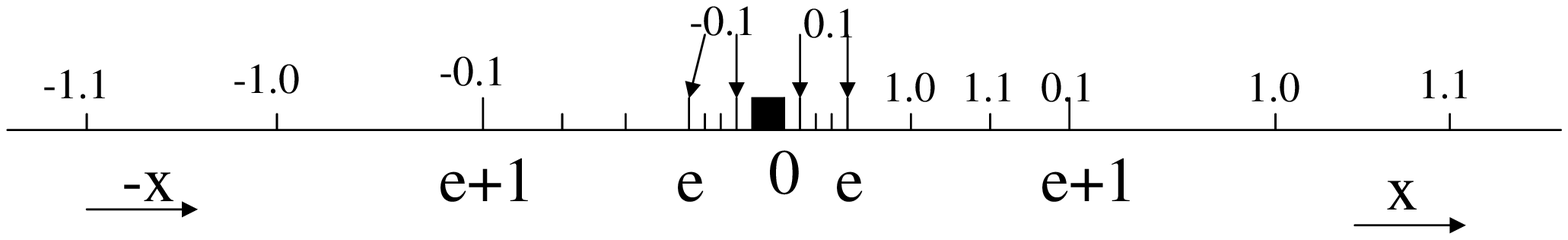}}\end{center}\caption{Locations
           for one dimensional $R_{n}$ space for $n=1$. Locations
           are shown by tick marks with jump locations labelled with
           $e,e\pm 1.$ The relative spacing of the ticks is shown
           on a locally flat background.} \label{2}\end{figure}

           Some of the properties are best seen by looking at a plot of the
           locations.  Figure \ref{2} shows such a plot for one
           dimension for $n=1.$  The locations
           are shown with their relative spacings on a locally flat background.
           This is the sheet of paper, computer screen etc. on
           which the figure is drawn.

           The figure is also drawn to show that for any scale factor $\underline{e}$ the
           spacing between adjacent points just to the left of the location $0.1,\underline{e}$
           is equal to the distance of the location $0.1,\underline{e}$ from the
           origin. It shows that regions of $2^{2n}-1$
           points with spacing $2^{n(2\underline{e}-1)}$ are separated by
           exponential jumps from adjacent regions of $2^{2n}-1$
           points with spacing $2^{n(2(\underline{e}\pm
           1)-1)}.$ Here $\underline{e}$ is any integer.

           The figure also shows the exponential crowding of the points
           towards the origin at $0.$ As a point of accumulation,
           $0$ is different from all other points.  It is unique
           in having no nearest neighbors.  This suggests that the
           origin is a space singularity or hole.

\begin{figure}[t!]\vspace*{1in}\begin{center}
          \resizebox{100pt}{100pt}{\includegraphics[300pt,100pt]
          [500pt,300pt]{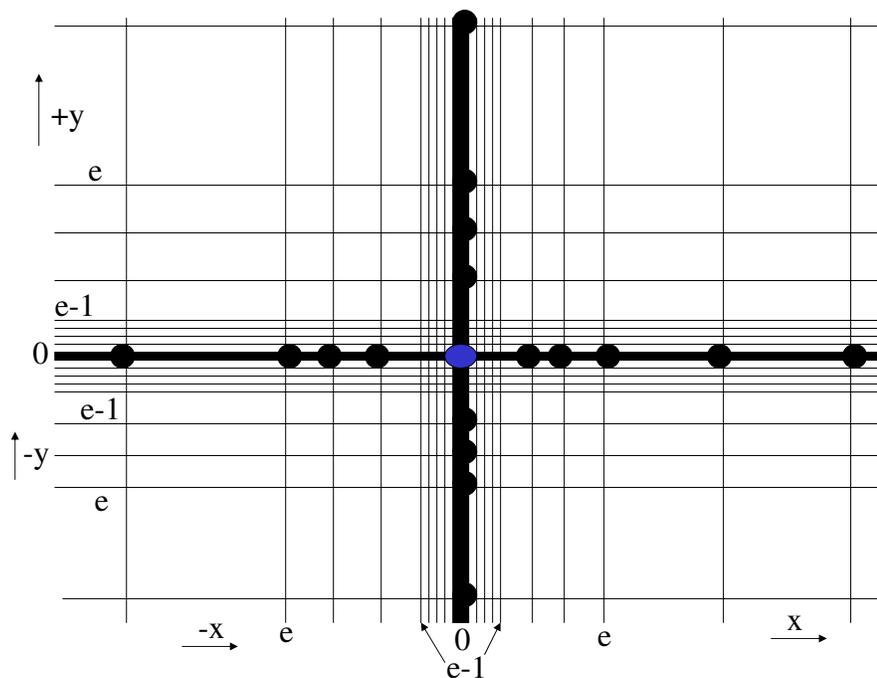}}\end{center}\caption{A two
          dimensional Cartesian plot of points in $R_{n}$ space
          for $n=1.$ The intersections of the lines denote space
          points and the intersections of thick lines with any
          line denote singularities.}\label{3}\end{figure}

           A cartesian coordinate plot in $2$ dimensional $R_{n}$ space for
           $n=1,$ Figure \ref{3}, shows these properties. In addition there are two types
          of singularities present. There is one two dimensional one at the
          origin and there are many one dimensional ones on the coordinate
          axes. In three dimensions there are three types of
          singularities, one three dimensional one at the origin,
          many two dimensional ones on the coordinate axes, and
          many one dimensional ones on the coordinate planes.

\begin{figure}[h]\vspace{5.2cm} \begin{center}
          \resizebox{100pt}{100pt}{\includegraphics[300pt,10pt]
          [520,240pt]{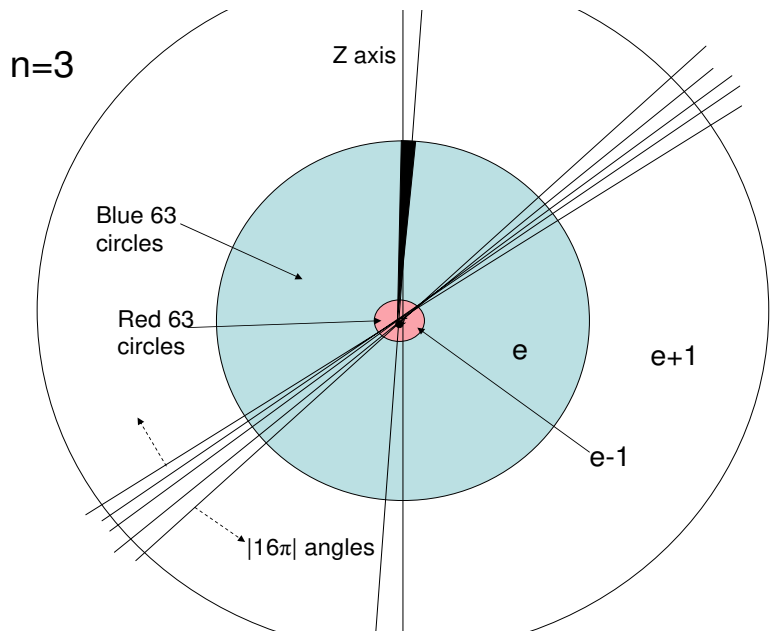}}\end{center}\caption{Polar
          coordinates for $R_{n}$ space for $n=3.$ The blue and
          red sections each contain $63$ concentric circles intersected
          by $|2^{n}\times 2\pi|=|16\pi|$ radii and an infinite number
          of radii crowded exponentially towards the $z$ axis.
          This is shown by the black wedge. There are an infinite number
          of circular sections also crowding exponentially to the origin.}
          \label{4}\end{figure}

          Figure \ref{4} shows a plot of polar coordinates
          in $2$ dimensional $R_{n}$ space for $n=3.$
           One sees
          that there is an infinite series of circular sections
          $R_{n,\underline{e}}$, one for each value of
          $\underline{e}.$  Each of these sections contains
          $2^{2n}-1 =63$ concentric circles intersected by an
          infinite number of radii.  However one is mainly
          interested in the radii with angular values in the
          numbers $R_{3,0}.$ There are $|2^{n}\times 2\pi|$ of
          these radii. These are shown in the figure. Note that
          the intersections of the radii and circles are the
          space locations.

          The radial depth of each section in the figure is
          smaller by a factor of $2^{2n}=64$ than the one outside
          it. Thus the red section in the figure is smaller
          by a factor of $64$ than the blue one.  Note that the circle
          outside the blue section is just the adjacent circle in the
          next section.  The full section would contain $63$ circles out
          to a radius $64$ times larger than that of the blue section. There is
          also an infinite sequence of exponentially shrinking
          radial sections, each with $63$ circles, towards the origin.

            The singularities consist of one at the origin and infinitely many
          others with one at the intersection of each circle with
          the $z$ axis.  This is shown by the black narrow wedge
          in the figure.

          Spherical coordinates in 3 dimensions are obtained by a $2\pi$
          azimuthal rotation of the two dimensional picture. Each
          $2$ dimensional circle becomes a $3$ dimensional sphere.
          There are an infinite number of azimuthal angles, but
          just as in the case of the polar angles one is mainly
          interested in  the $|2^{n}\times 2\pi|$ azimuthal
          angles for the scale factor $\underline{e}=0$
          for the angular coordinates. The reason is that these
          cover the azimuth angular range of $0$ to $2\pi$ in increments of
          $2^{-n}.$ It follows that each concentric sphere is
          intersected by $|2^{n}\times 2\pi|\times
          |2^{n}\times\pi|$ radius vectors, one for each pair of
          angles.

          An infinite number of azimuthal angles concentrate
          exponentially near the $\phi =0$ plane. The
          singularities consist of one at the origin, a line of
          $2$ dimensional singularities along the $z$ axis, and a
          plane of $1$ dimensional singularities for $\phi =0$.
          Note that the $2$ and $1$ dimensional singularities are one
          sided. Also the differences
          between point locations for Cartesian and spherical
          systems and between those in the figures and in the usual $R$ space
          become unobservable as $n$ gets large.

          A transformation can be defined on $R_{n}$ space
          that is based on the ordering of the numbers in $R_{n}.$
          Let $x_{\underline{s},2n\underline{e}}$ be a one
          dimensional component of a point in $R_{n}$ space
          relative to some coordinate system. Here $\underline{s}$
          is a length $2n$ binary string with the "binal point"
          separating the $nth$ and $n+1st$ elements of
          $\underline{s}.$ A location in $3$ dimensional space is
          described by a triple of these components.

          The basic ordering on the numbers in $R_{n}$ is defined
          by  \cite{BenTCTPMTEC}
        \begin{equation}\label{orderadd}f_{<}(\underline{s}
        ,2n\underline{e})=\left\{\begin{array}{ll}(\underline{s}.
        ,2n\underline{e})+(\underline{0}1,2n\underline{e}) &
        \mbox{ if }\underline{s}\neq \underline{1} \\
        (\underline{0}1,2n(\underline{e}+ 1))& \mbox{ if }\underline{s}
       =\underline{1}\end{array}\right .\end{equation} Here $\underline{1}$ is
       the constant $1$ string and $\underline{0}1$
       is the string of all $0s$ except the last or $2nth$ element
       which is a $1$. Note that the zero strings $\underline{0},2n\underline{e}$ are not
       in the domain or range of $f_{<}.$

       The above definition is for positive $\underline{s}.$ for
       negative $-\underline{s}$ the corresponding definition is
       \begin{equation}\label{orderaddneg}f_{<}(-\underline{s},2n\underline{e})
        =\left\{\begin{array}{ll}(-\underline{s},2n\underline{e})+
        (\underline{0}1,2n\underline{e}) & \mbox{ if }\underline{s}
        \neq \underline{0}1 \\ (-\underline{1},2n(\underline{e}- 1))& \mbox{ if }
        \underline{s}=\underline{0}1\end{array}\right.\end{equation} Note
        that, in line with the usual order properties of integers,
        $f_{<}$ applied to positive numbers increases their
        magnitude.  It decreases the magnitude of negative
        numbers.

        The inverse operation $f_{>}=f^{-1}_{<}$ can be defined from $f_{<}$ by
       \begin{equation}\label{invorderadd}f_{>}
       (f_{<}(\underline{s},2n
       \underline{e})= (\underline{s},2n\underline{e}).
       \end{equation} The definition of $f_{<}$ on negative $\underline{s}$ is
       related to the inverse $f_{>}$ by
       \begin{equation}\label{orderneg}f_{<}(-\underline{s},2n\underline{e})
       =-(f_{>}(\underline{s},2n\underline{e})).
       \end{equation} This says that moving
       along with the ordering on negative numbers is equivalent
       to moving opposite to the
       ordering on the positive numbers and changing the sign.

       These definitions can be used to define a transformation
       $F_{<}$ on $R_{n}$ space by
       \begin{equation}\label{Fdef} F_{<}(x_{\underline{s},2n\underline{e}})=
       x_{f_{<}(\underline{s},2n\underline{e})}. \end{equation}
       $F_{<}$ has an inverse, $F_{>}=F_{<}^{-1}$ defined in the
       obvious way.

\begin{figure}[t!]\vspace{-1cm} \begin{center}
       \resizebox{150pt}{140pt}{\includegraphics[250pt,180pt]
       [470pt,400pt]{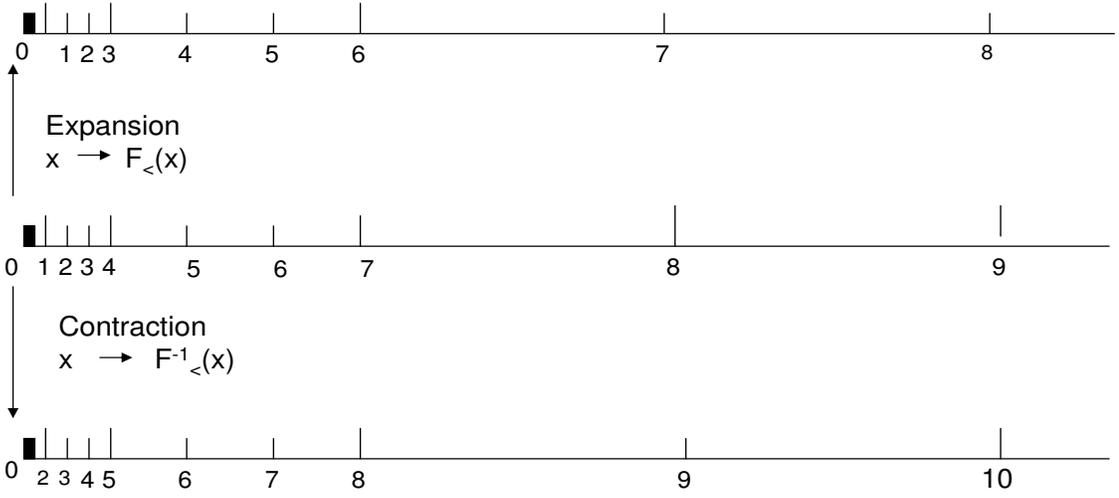}}\vspace{1cm}\caption{Action of $F_{<}$
       and its inverse on radii and positive Cartesian coordinate
       locations. Points are distinguished by numbers under the
       tick marks.}\label{5}
       \end{center}\end{figure}

       The action of $F_{<}$ is shown in Figure \ref{5} for
       positive coordinate values, such as radii or positive
       Cartesian locations. The numbers under the locations serve
       as distinguishing labels.  The figure shows that the action
       of $F_{<}$ and of its inverse correspond to expansion and
       contraction of $R_{n}$ space.  This is shown by the motion
       of points away from the origin under $F_{<}$ and towards
       the origin under $F_{<}^{-1}.$ Scale invariance is shown by
       the same locations of the tick marks on each of the three
       lines, relative to the locally flat background on which the
       figure is drawn.

       The origin is unique in that it is the only point whose
       location is unchanged under expansions or contractions.  In
       this sense it, or an arbitrarily small neighborhood around
       the origin, can be regarded as a source or sink of space
       points for expansions or contractions respectively.

       \section{Inflationary Cosmology on $R_{n}$
       Space}\label{ICRS}

       Here the description of inflationary cosmology is limited to
       showing that $R_{n}$ space is a suitable framework for
       describing elementary aspects of the big bang, inflation,
       and the redshift or Hubble expansion after the inflation.
       No physical theory will be given to support the dynamical
       model used.

       In usual $R$ space time, inflationary cosmology is based on
       the Friedmann Robertson Walker (FRW) equation
       \cite{Brandenberger} for the distance
       element\begin{equation}\label{FRW}
       ds^{2}=c^{2}dt^{2}-A(t)[\frac{dr^{2}}{\sqrt{1-kr^{2}}}+r^{2}
       (d^{2}\theta +\sin^{2}{\theta}d^{2}\phi)].\end{equation}
       Here $k=0$ for asymptotically flat space will be chosen.

       A space singularity at $t=0$, that contains all space and
       explodes at $t=0$, corresponds to the Big Bang. An inflationary era,  with
       exponential expansion given by \begin{equation}\label{ExpA}
       A(t)=e^{Ht},\end{equation} begins almost immediately. Here
       $H$ is the inflationary Hubble constant.  During this
       period the universe expands by a factor of at least $10^{50}$
       in about $10^{-23}$ seconds \cite{Guthstein}. Inflation
       ends with a sudden drop in $$H(t)=\dot{A}/A(t)$$ to a
       present value of $H=71\pm 7\; km/sec/mpc$ \cite{Freedman}.
       This gives the redshift.

       The success of inflation is based on the solution of
       several problems in the standard big bang cosmology.
       Included is the horizon problem, the flatness problem, and
       the large scale galactic clumping problem
       \cite{Brandenberger,Guthstein,Watson,Guth,Linde}. These
       problems are based on the need to explain the large scale
       uniformity of the universe, the asymptotic flatness of
       space, and the large scale clumping of galaxies. Details
       are given in the literature referenced above. Here it
       is sufficient to note that these problems are solved by a
       brief period of superluminal  exponential expansion of both the
       observable universe and the horizon distance.

       Here it will be seen that a constant iteration rate of
       $F_{<},$ coupled with a time dependence of the scale factor
       $\underline{e}$ and $n,$ gives a superluminal expansion
       velocity to part of $R_{n}$ space. This is referred to here
       as inflation on $R_{n}$ space.

       One begins with  the Big Bang. This sets the cosmological
       time at $t=0$ and a location that is a
       $3d$ singularity or coordinate origin in $R_{n}$ space.  Only the
       origin appears as space is isotropic. This breaks the scale
       invariance as all of space is concentrated in a small
       region around the origin.

       The method used here is to limit the scale factor
       $\underline{e}$ in $R_{n}$ space to be less than some
       time dependent maximum $\underline{e}_{max}(t).$  Then at time $t$ $R_{n}$ space
       is limited to points $x_{\underline{s},2n\underline{e}}$
       where $\underline{e}\leq \underline{e}_{max}(t).$

       The initial conditions at $t=0$ are given by
       \begin{eqnarray} \underline{e}_{max}(0)=\underline{e}_{0} \nonumber \\
       n=n_{0}\end{eqnarray} where $\underline{e}_{0}$ is
       negative or possibly $0$.  Note that the restriction on
       $\underline{e}$ does not apply to the numbers in $R_{n}.$
       It applies to the locations in $R_{n}$ space only.  At
       $t=0$ all $R_{n_{0}}$ space is confined to a region of radius
       $2^{n_{0}(\underline{e}_{0}+1)}$ around the origin.

       The dynamics of inflation is described here by repeated
       iteration of $F_{<}$ on the radial components of space
       only. The iteration rate will be taken here to be constant
       with rate constant $\beta.$ After time $t$ $|\beta t|$
       denotes the number of iterations with $F_{<}^{|\beta t|}$
       the corresponding map on $R_{n_{0}}$ space.

       The constant iteration rate is chosen here because it is
       simple and is sufficient to give an inflation to
       superluminal velocities in $R_{n}$ space. If desired one
       may choose more complex time dependent iteration rates for
       investigation.  However the simple constant iteration rate
       assumption is sufficient for illustrative purposes.

       The important point here is that a constant iteration rate
       of $F_{<}$ gives an exponential expansion rate of space.
       After $m(2^{2n_{0}}-1)$ steps, distances expand by a factor
       of $2^{2n_{0}m}$.  This can be seen from Fig. \ref{5}.
       After a time $t$ with $|\beta t|=m(2^{2n_{0}}-1),$
       distances expand by a factor of $2^{2n_{0}m}.$  This is the
       source of inflation in $R_{n_{0}}$ space.

       A space expansion factor corresponding to $A(t)$ in the FRW
       equation, Eq.  \ref{FRW}, can be defined here.  To this end let
       $D(x_{2},x_{1},0)$ be the distance between points
       $x_{2},x_{1}$ at step or time $0$. The distance
       $D(x_{2},x_{1},j)$ at step $j$ is related to
       that at step $0$ by
       \begin{equation}\label{Aj}D(x_{2},x_{1},j)=A(j)D(x_{2},x_{1},0).
       \end{equation} where $A(j)=A(|\beta t|)$ is equivalent to
       $A(t)$ in the FRW equation.

       However, unlike the FRW $A(t),$ $A(j)$ depends on $x_{2}$
       and $x_{1}.$  This is a consequence of the structure of
       $R_{n_{0}}$ space as a collection of scale sections
       $R_{n_{0},\underline{e}}$ over an infinite number of
       $\underline{e}$ values. Also the fact that the distance
       metric $D$ is a map into the numbers $R_{n_{0}}$ with
       their arithmetic, influences the values of $A(j).$ For
       example, if $x_{2}$ is in $R_{n_{0},\underline{e}_{2}}$ and
       $x_{1}$ is in $R_{n_{0},\underline{e}_{1}}$ and
       $\underline{e}_{2}-\underline{e}_{1}\geq 2$, then
       $D(x_{2},x_{1},j)=D(x_{2},0,j)$ which is the distance of
       $x_{2}$ from the origin. These relations hold for any
       $n$,not just $n_{0}$.

       It is useful to note that $A(j)$ can be
       expressed as an iterative product of factors $a(j+1,j)$
       according to \begin{equation}\label{defaj}
       A(j+1)=a(j+1,j)A(j).\end{equation} Based on this one has
       \begin{equation}\label{Djj1}
       D(x_{2},x_{1},j+1)=a(j+1,j)D(x_{2},x_{1},j).
       \end{equation}

       The expansion velocity $V(x_{2},x_{1},j)$ of the distance
       between $x_{2},x_{1}$ at step $j$ is given by
       \begin{equation}\label{Velj}
       V(x_{2},x_{1},j)=\frac{a(j+1,j)-1}{\Delta
       j}D(x_{2},x_{1},j).\end{equation}
       The equivalent equations for the time $t$
       are obtained by setting $j=|\beta t|,\;
       j+1=|\beta(t+\beta^{-1})|$ and replacing $\Delta j=1$ by
       $\Delta t=\beta^{-1}.$

       The values of $V(x_{2},x_{1},j)$ vary widely depending on
       the relationship between $x_{2}$ and $x_{1}.$ If $x_{1}=0$
       then $V(x_{2},x_{1},j)$ is constant for
        $2^{2n_{0}}-1$ steps, then an increase by a factor of
        $2^{2n_{0}}$ and then constant at the higher level. If
        $x_{2},x_{1}$ are on the same radius vector in the same
        scale section $R_{n_{0},\underline{e}},$ then
        $V(x_{2},x_{1},j)=0$ for some $j$ and very large for other
        $j$.

        These features can be seen by examining the points in
        Fig. \ref{4}. The expansion velocity of point
        $4$, (centerline) from the origin  is constant for $3$ steps
        until it reaches the location of point $7$. Then it jumps
        by a factor of $4$ and is constant at the higher value for
        another $3$ steps.  The expansion velocity of the distance
        between points $4$ and $5$ is $0$ for $3$ steps until the
        point $5$ reaches the location $7$ and $4$ reaches location $6$.
        Then the velocity jumps  to high values and back down as
        point $6$ moves to $7$ and $7$ moves to $8.$

        The expansion velocity, averaged over $2^{2n_{0}}-1$
        steps, increases exponentially.  To see this define the
        averaged velocity by \begin{equation}\label{Vave}
        \langle V(x_{2},x_{1},m)\rangle
        =\frac{A(2^{2n_{0}}-1)-1}{2^{2n_{0}}-1}D(x_{2},x_{1},j_{m})
        \end{equation}where $j_{m}=m(2^{2n_{0}}-1)$ and
        $A(2^{2n_{0}}-1)=\prod_{q=j_{m}}^{j_{m}+2^{2n_{0}}-1}a(q+1,q)$. Since
        $A(2^{2n_{0}}-1)=2^{2n_{0}}$ independent of $j_{m}$, or of
        any $j$ value, one has \begin{equation}\label{VaveD}
        \langle V(x_{2},x_{1},m)\rangle = D(x_{2},x_{1},j_{m}).
        \end{equation} This gives  \begin{equation}\label{Vaveexp}
        \langle V(x_{2},x_{1},m+1)\rangle
        =2^{2n_{0}}  \langle V(x_{2},x_{1},m)\rangle
        \end{equation}In terms of the time one has \begin{equation}
        \label{Vavet}\langle V(x_{2},x_{1},t_{m+1})\rangle
        =2^{2n_{0}}\langle V(x_{2},x_{1},t_{m})\rangle
        \end{equation} where $|\beta t_{m}|=m(2^{2n_{0}}-1).$

        It follows that at some time $t$ the average expansion
        velocity will be greater than $c$, the velocity of light.
        It is of interest  to determine the relation between the
        duration $t_{I}$, of inflation, which here is the era of the
        exponential expansion described above, and the number of
        superluminal locations at time $t_{I}.$ To this end recall
        that $R_{n_{0}}$ space consists of an infinite sequence of
        concentric spherical shells, $R_{n_{0},\underline{e}}$
        each containing $(2^{2n_{0}}-1)|2^{n_{0}}\pi||2^{n_{0}+1}\pi|$
        points. At time $0$ the band
        $R_{n_{0},[\underline{e}_{0}-\Delta,\underline{e}_{0}]}$
        of $\Delta$ scale sections $R_{n_{0},\underline{e}}$ with
        $\underline{e}_{0}-\Delta < \underline{e}\leq
        \underline{e}_{0}$ has $N_{\Delta}$ locations where
       \begin{equation}\label{NDelta}
        N_{\Delta}=\Delta (2^{2n_{0}}-1)|2^{n_{0}}\pi||2^{n_{0}+1}\pi|
        \end{equation}

        The least number of
        steps $j_{I}=m_{I}(2^{2n_{0}}-1)$ to give all locations in
        $R_{n_{0},[\underline{e}_{0}-\Delta,\underline{e}_{0}]}$
        velocities $>c$ from the big bang origin is obtained from the velocity inequality
        $\beta d\times 2^{n_{0}(2(\underline{e}_{0}-\Delta +m_{I})+1)}>c.$
        The result is
        \begin{equation}\label{jImI}
        m_{I}\geq -\underline{e}_{0}+\Delta +\frac{\log_{2}{(\beta
        d)^{-1}c}}{2n_{0}}+\frac{1}{2}.
        \end{equation} Here $d$ is a unit distance parameter. It is
        needed to make $(\beta d)^{-1}c$ a ratio of velocities and a
        pure number.
        Note that the least time $t_{I}$ is given by
        $j_{I}=|\beta t_{I}|.$

        It turns out that one can determine from the above
        relations the value of $\Delta$ that maximizes inflation
        efficiency in that it maximizes the value of $N_{\Delta}$
        for given $j_{I}$ or $t_{I}.$  The maximally efficient
        $\Delta$ is given by \begin{equation}\label{maxdelta}
        \Delta = (-\underline{e}_{0}
        +\frac{\log_{2}{(\beta d)^{-1}c}}{2n_{0}}+\frac{1}{2}).
        \end{equation} Combining this with Eq. \ref{jImI} gives
        \begin{equation}\label{delmI}\Delta = \frac{m_{I}}{2}
        \end{equation}

        So far one sees that a constant iteration rate $\beta$ of
        $F_{<}$ gives an exponentially expanding space out from
        the origin. The above equation give the relations  between the initial values
        of $\underline{e}_{0},n_{0}$ and the minimal time needed
        to accelerate all points in the band
        $R_{n_{0},[\underline{e}_{0}-\Delta,\underline{e}_{0}]}$
        to superluminal velocities away from the origin. At time
        $t_{I}$ $R_{n_{0}}$ space is a finite sphere with radius
        $2^{n_{0}(2(\underline{e}_{0}+m_{I})+1)}.$ It contains an
        infinite number of scale sections
        $R_{n_{0},\underline{e}}$ where $\underline{e}\leq
        \underline{e}_{0}+m_{I}.$ The space does not look anything
        like the usual $R$ space due to the presence of
        exponential jumps at the boundaries of each scale section.

        Another problem is that the model presented so far gives no
        way to stop inflation.  This is clearly necessary because
        inflation took place only during the earliest part of the
        $14.4$ billion years of cosmic evolution,

        Both these problems are taken care of here by making $n$
        time dependent.  That is $n=n(j)=n(|\beta t|).$  This is a
        much more fundamental time dependence than that of the
        scale factor.  The reason is that it means that the numbers $R_{n(t)}$
        and $C_{n(t)}$, and all physical theories based on these numbers,
        become time dependent.  Since there is no empirical
        evidence for such a time dependence one must ensure that
        it is completely invisible at the present time.

        The model chosen here to solve these problems and stop
        inflation is a time dependence given by \begin{equation}\label{ntimedep}
        \begin{array}{ll}n(t)=n_{0} & \mbox{ if
        $0\leq t\leq t_{I}$}\\n(t)\rightarrow n_{I}>>n_{0} & \mbox{ if
        $t\geq t_{I}+\delta.$}\end{array} \end{equation} For
        simplicity in the following $\delta=0$ will be assumed.
        This makes the time dependence of $n$ a simple step
        function of the time with $n(t)=n_{0}$ if $t\leq t_{I}$
        and $n(t)=n_{I}$ if $t\geq t_{I}.$

        The main conditions on $n_{I}$ are that it should be such
        that inflation stops at $t=t_{I}$ and that the band
        $R_{n_{0},[\underline{e}_{0}-\Delta,\underline{e}_{0}]}$
        is contained in the scale section $R_{n_{I},0}$ of
        $R_{n_{I}}$ space. The stopping of inflation follows from
        the drop of the expansion velocity $ V$ from  values $>c$ to
        \begin{equation}\label{VnI}V=\beta d 2^{-n_{I}}
        \end{equation} at time $t_{I}.$  This is based on a
        separation distance of $2^{-n_{I}}d$ between adjacent
        radial points in $R_{n_{I}}$ space and a constant
        iteration rate $\beta$, of $F_{<}$.

        The stopping of inflation can also be seen quite clearly
        by replacing the exact expression for the expansion factor
        $A(t)=A(|\beta t|)$ by an exponential approximation
       \begin{equation}\label{Aexpn} A(t)=2^{H_{n}|\beta t|}.
        \end{equation}  The expansion coefficient
        $H_{n}=2n/(2^{2n}-1)$ is such that this approximation
        equals the exact expression at each time $t$ where $|\beta
        t|=m(2^{2n}-1)$ for some $m$.  From this viewpoint
        inflation ends when $H_{n}$ drops from
        $2n_{0}/(2^{2n_{0}}-1)$ to $2n_{I}/(2^{2n_{I}}-1).$

         The containment requirement gives a lower limit for
        $n_{I}:$ \begin{equation}\label{nI}
        n_{I}\geq 2n_{0}\Delta+\log_{2}{(\beta d)^{-1}c}+2n_{0}.
        \end{equation} This requirement on $n_{I}$ is sufficient
        to satisfy the condition that the scale section $R_{n_{I},0}$ be
        experimentally indistinguishable from the usual $R$ space.
        This means that no exponential jumps should be present
        over the range of experimentally accessible space and the
        discreteness should be too fine to be observed. These
        requirements translate to conditions on $n_{I}$ in that
        the scale section contains no exponential
        jumps, except at its borders, and it extends from
        $2^{-n_{I}}$ to $2^{n_{I}}$ in $2^{2n_{I}}$ steps of width
        $2^{-n_{I}}.$ Note also that the scale section $R_{n_{I},0}$
        has many more space points than the band
        $R_{n_{0},[\underline{e}_{0}-\Delta,\underline{e}_{0}]}.$

        \subsection{Numerical Examples}\label{NE}

        It is quite useful to give some numerical examples to see
        how the model described here works.  The following table gives a
        summary of some examples for different initial conditions
        as different values of $\underline{e}_{0}$ and $n_{0}.$
        The values of $\log_{2}{(\beta d)^{_1}c}=20$ and $\beta =10^{-6}
        sec^{-1}$ are chosen arbitrarily and are the same for all table
        entries. From these values and the value for $c$ one obtains that
        $d=0.03\; cm.$ $$\begin{array}{c}\mbox{ Table of
          Examples: }\\\begin{array}{|c|c||c|c|c|c|c|c|c|}\hline
           \underline{e}_{0} & n_{0}  & m_{I} & \Delta & R_{IO}&
           N_{\Delta} &  n_{I}  & R_{FO} & \tau_{I} \\ \hline -3 &
           20 & 8 & 4 & 10^{-30} & 7\times 10^{25} & 220 & 10^{66}
           & 90 days \\ \hline -2 & 20 & 6 & 3 & 10^{-18} &
           7\times 10^{25} & 180 & 10^{54} & 80 days \\ \hline
            -10 & 10 & 23 & 11 & 10^{-57} & 2\times 10^{14} &  270 & 10^{81} & 23 sec
            \\ \hline -5& 10 & 13 &  6 & 10^{-27} & 1\times 10^{14} & 160 & 10^{51} & 13 sec
             \\ \hline -10 & 5 & 25 & 12 & 10^{-28} & 2\times 10^{8} &
            150 & 10^{46} & 0.025 sec \\ \hline
            -5 & 5 & 15  & 7 & 10^{-14} & 1\times 10^{8} &  100 &
            10^{31} & 0.015 sec \\ \hline -2 & 5 & 9 & 5 & 10^{-5} &
            1\times 10^{8} & 75 & 10^{22} & 0.009 sec \\ \hline 0 & 5 & 5
            & 2 & 32 & 4\times 10^{7} &
            55 & 10^{17} & 0.005\; sec \\ \hline -10 & 3 & 28 & 14 &
            10^{-17} & 1\times 10^{6} & 110 & 10^{33} & 0.002 sec
            \\ \hline -5 & 3 & 18 & 9 & 10^{-8} & 7\times 10^{5} &
            80 & 10^{24} & 0.001 sec \\ \hline -2 & 3 & 12 & 6 &
            0.002 & 5\times 10^{5} & 62 & 10^{19} & 8\times
            10^{-4} sec \\ \hline\end{array}\end{array}$$\\

        The table entries are based on values of other quantities needed to give
        superluminal expansion velocities to all points in the
        band $R_{n_{0},[\underline{e}_{0}-\Delta,
        \underline{e}_{0}]}.$  $m_{I}$ is given by Eqs. \ref{jImI}
        and \ref{maxdelta}.
        The value of $\Delta$ is given as the integer roundoff
        of $\Delta =m_{I}/2.$ $N_{\Delta}$ is the number of points
        in the band, and $R_{IO}=2^{n_{0}(2\underline{e}_{0}+1)}$
        and $R_{FO}=2^{n_{0}(2(\underline{e}_{0}+m_{I})+1)}$ are
        the outer radii of the band at the outset and at the time
        $\tau_{I}$ when inflation stops after $m_{I}$ increases in
        scale section  values. The initial and final values for the
        innermost radius of the bands are given by
        $R_{II}=2^{n_{0}(2(\underline{e}_{0}-\Delta)+1)}$ and
        $R_{FI}=2^{n_{0}(2(\underline{e}_{0}-\Delta +m_{I})+1)}.$
        The value of $n_{I}$ from Eq. \ref{nI} is the
        minimal value needed to satisfy the containment condition.

        The table entries show that the value of $m_{I}$ is quite
        sensitive to the value of $\underline{e}_{0}.$ This is to
        be expected because most of the time is spent bringing the
        scale values of the band from negative values up to values
       $ \geq 0.$  This can be seen from the fact that $m_{I}$ is
       only slightly larger than $\Delta +|\underline{e}_{0}|.$

       The values of $R_{FO}d$ should be regarded as the horizon
       radius of the universe after inflation. Since this is the radius
       of $R_{n_{0}}$ space it is an upper bound to the radius of the
       observable universe at any time, including the end of inflation.
       Thus the horizon problem is avoided here. Note also that for several
       initial conditions the values of $R_{FO}d$ are much greater than the
       radius of the present observable universe ($\sim
       10^{28}cm$). Also inflation factors $R_{FO}/R_{IO}$
       range from $10^{138}$ down to $10^{15}.$

       The times of inflation range from $90$ days to $<10^{-3}$ secs
       for the table entries. These show that  inflation in the model considered here is
       quite leisurely compared to the model of
       Guth \cite{Guth} and Linde \cite{Linde}. This model gives an
       inflation factor of
       about $10^{50}$ in $10^{-32}$ second \cite{Guthstein}, which is
       much faster. Whether this difference is essential and
       inflation has to be so fast depends on future work.

       Here if desired, inflation can be speeded up by increasing
       the value of $\beta$ or requiring that the iteration rate
       of $F_{<}$ be exponentially dependent on time.  The point
       of the discussion here is to show that, with $R_{n}$ space as
       a framework, this is not needed to achieve superluminal
       velocities.

        The table entries also show that
        the value of $N_{\Delta}$ is quite sensitive to
        the value of $n_{0}$. Large values of $N_{\Delta}$ are
        desirable since this is the number of superluminal points
        that may act like seeds for regions in $R_{n_{I},0}$ space
        that start out at the end of inflation with no causal
        connection to the past.  Here one notes that the number of
        points in $R_{n_{I},0}$ space is much greater than the
        number of points of $R_{n_{0}}$ space that correspond to
        postinflationary seeds in $R_{n_{I},0}$ space.

        Since physical systems at these seed locations were
        causally connected at the outset, they bring with them
        information and physical properties based on the past
        connections.  This includes any uniformity or
        nonuniformity of properties. Since causal connections to the past
        are broken by inflation for the superluminal points, the
        systems at these locations can interact with others
        created at the nearby new points of $R_{n_{I},0}$ space
        and evolve independently of systems around other seed
        regions.

        The table entries also show that the values of $n_{I}$ are
        sensitive to the values of $n_{0}$.  This is mainly a
        consequence of the containment requirement and the
        resulting lower bound on $n_{I}$, Eq. \ref{nI}, used to
        compute the values. As noted before, this condition is
        quite strict as one really requires that the $R_{n(t),0}$
        section of $R_{n(t)}$ space be experimentally indistinguishable
        from $R$ space at the present time with $n(t)$ at its present
        value which  is greater than $n_{I}.$

        One way to express the requirement of experimental
        indistinguishability is to require that $n(t)$ be such
        that the range of the $R_{n(t),0}$ section extend from the
        Planck length to the radius of the universe at the present
        time. This is equivalent to the condition \begin{equation}\label{nt}
        2^{2n(t)}\geq\frac{\mbox{Universe radius at present time}}{\mbox{
        Planck length}}\end{equation} or $n(t)\geq 102.$  The
        table shows that this condition is already satisfied for
        many entries.  The other entries with lower values of
        $n_{I}$ do not cause a problem as the Hubble expansion
        after the end of inflation can easily take care of the
        difference.

        \subsection{Hubble Expansion and the Redshift}\label{HER}
        After inflation ends the universe continues to expand and
        is doing so at present.  This is shown by the redshift of
        light from distant galaxies with an expansion parameter or
        Hubble constant of $71\pm 7\; km/sec/mpc$ \cite{Freedman}.
        (A megaparsec is about $3.26\times 10^{6}$ light years.)
        Whether the universe expansion is slowing down or speeding
        up, due to the possible presence of dark energy, is a
        matter of much debate at present \cite{Bahcall}.

        Here the Hubble expansion is accounted for by a
        continuing increase in $n(t)$ after inflation ends. The
        model used is based on replacing $a(j+1,j)$ in Eq.
        \ref{defaj} by \begin{equation}\label{defbj}
        b(j+1,j)=a(j+1,j)e^{\epsilon\Delta n(j)}.\end{equation}
        Here $\Delta n(j) =n(j+1)-n(j)$ with $j=|\beta t|$ and
        $j+1=|\beta(t+\beta^{-1})|.$  The corresponding change in
        the $A(j)$ factors replaces $A(j^{\prime},j)=
        \prod_{q=j}^{j^{\prime}}a(q+1,q)A(j)$ by
        $B(j^{\prime},j)=A(j^{\prime},j)e^{\epsilon
        (n(j^{\prime})-n(j))}.$ This holds for all $j\geq j_{I}$ and
        $j^{\prime}\geq j.$

        Here to keep things simple the replacement is made after
        the end of inflation.  However extending the use of the
        factor $e^{\epsilon\Delta n}$ to all time before and after
        inflation is straightforward. For times up to $t_{I}$ the
        factor contributes nothing. For the change at $t_{I}$ of
        $n$ from $n_{0}$ to $n_{I}$ an additional factor of
        $e^{\epsilon (n_{I}-n_{0})}$ is present. It will be seen
        that $\epsilon$ is sufficiently small so that this factor
        also contributes almost nothing as  $e^{\epsilon (n_{I}-n_{0})}\sim 1.$

        Again it must be emphasized that this model is chosen to
        show the suitability of $R_{n}$ space as a framework for
        inflationary cosmology. No physical theory is given to
        support the choice.  However a very naive justification for the
        factor $e^{\epsilon\Delta n}$ can be provided by
        considering length $2n$ binary strings as basic elements,
        such as a field of these strings.  Excited states of these
        strings correspond to values of physical parameters or
        just numbers depending on which parameter is excited. If
        one assumes that the space occupied by the strings depends
        on $n,$ then increasing $n$ pushes space apart and
        increases the distance between points. In this view the
        value of $\epsilon$ describes the rigidity of space or the
        resistance to being expanded.

        As noted the time dependence of $n(t)$ induces a time
        dependence of the numbers  $R_{n(t)}$ and on
        $R_{n(t),0}$ space. (From now on the $\underline{e}=0$
        space section $R_{n(t),0}$ is referred to as $R_{n(t),0}$
        space.) For the numbers the change $n\rightarrow n+1$
        induces the change
        $$R_{n}=\{\underline{s}_{[1,n]}.\underline{s}_{[n+1,2n]},2n\underline{e}\}
        \rightarrow\{\underline{s}_{[1,n+1]}.
        \underline{s}_{[n+2,2n+2]},2(n+1)n\underline{e}\}=R_{n+1}.$$

        The corresponding change $R_{n,0} \rightarrow
        R_{n+1,0}$ space maps each component location
        $x_{\underline{s}_{[1,2n]},0}$ into
        $x_{\underline{0s0}_{[1,2n+2]},0}.$  Note the addition of
        one leading and one trailing $0$. From Eq. \ref{NDelta} one sees that there are about $16$
        times as many points in $R_{n+1,0}$ space as in $R_{n,0}$
        space. If $\Delta$ is the distance between two points in
        $R_{n,0}$ space the distance between the corresponding two
        points in $R_{n+1,0}$ space is $\Delta e^{\epsilon}.$ The
        distance between adjacent points on a radius vector (Fig. \ref{5})
        in $R_{n+1,0}$ space is $e^{\epsilon}/4$ times the corresponding
        distance in $R_{n,0}$ space.  For adjacent points on the
        same sphere the corresponding factor is $e^{\epsilon}/2$
        for each angular coordinate.

        The specific time dependent model for $n(t)$ chosen here
        is that for a constant expansion rate: \begin{equation}\label{modelnt}
        n(t) =|\gamma (t-t_{I})|+n_{I}.\end{equation} Here $\gamma$
        is the rate constant and $n(t+\gamma^{-1})=n(t)+1.$ The
        Hubble expansion parameter is obtained as the discrete
        form of $\dot{A}(t)/A(t).$ For $t\geq t_{I}$ one gets
        \begin{equation}\label{Hge}H=\gamma\epsilon.
        \end{equation} The experimental value of $H=71\pm 7\;
        km/sec/mpc$ \cite{Freedman} gives \begin{equation}\label{geexp}
        \gamma \epsilon=7.3\pm 6\times 10^{-11}\; year^{-1}.\end{equation}

        One experimental prediction that follows from this is that
        the recession velocity of distant objects is a step
        function of the distance.  This follows from the
        discreteness of the values of $n(t).$
        If $D$ is the distance of a galaxy, then
        the recession velocity is given by $cL\epsilon$ where
        $L=|D\gamma /c|.$  Clearly $L=0$ if $D<\gamma /c.$

        The empirical data plotted as recession velocity against
        distance \cite{Freedman} do not show evidence of a step function. However
        there is considerable scatter among the points.  The empirical
        data can be used to put an upper limit on $\epsilon$ in that it
        must be sufficiently small so steps are not observed
        within experimental error. This puts a corresponding lower
        limit on $\gamma$, Eq. \ref{geexp}.  A simple examination of the  data
        plot given by Freedman \cite{Freedman} shows that a  unit step
        increase of $n(t)$ every $30-60$ million years would not be
        observable. This gives an upper limit of $\epsilon\leq
        2-4\times 10^{-3}.$

        \section{Summary and Discussion}
        A space and time was described that inherits the properties of
        numbers corresponding to  $2n$ figure outputs of measurements of
        physical quantities of infinite range. The space, $R_{n}$ space, can
        be described as an infinite sequence of spherical scale sections
        $R_{n,\underline{e}}$ where $\underline{e}$ is any
        integer. Each section has the same number of points but
        the size of each region increases or decreases exponentially with
        increasing or decreasing $\underline{e}.$ The center is a
        point of accumulation of the sections and is a space
        singularity.

        A model of inflationary cosmology is described that is based on
        these properties of $R_{n}$ space and time. The dynamics
        is described by iteration of a basic order preserving transformation
        $F_{<}$ and a time dependence of $n=n(t)$. The properties
        of $R_{n}$ space are such that iteration of $F_{<}$ at a
        constant rate corresponds to an exponential expansion of
        space. The localization of space at the origin is done by
        requiring that at cosmological time $t\sim 0,$ $R_{n}$ space
        is limited to scale sections $R_{n,\underline{e}}$ with
        $\underline{e}\leq\underline{e}_{0}$ a negative integer.
        Also the initial value, $n_{0},$ of $n$ is a small positive
        integer.

        Based on a constant iteration rate of $F_{<}$, inflation occurs
        naturally until the outermost $\Delta$ scale sections are
        expanding away from the center at velocities $>c$, the
        velocity of light.  At this time $t_{I}$
        the value of $n=n(t)$ is increased from $n_{0}$ to
        $n_{I}>n_{0}.$ Inflation stops because the increase in $n$ drops the expansion
        velocities from values $>c$ to $2^{-n_{I}}\beta d.$

        The value of $n_{I}$ is determined by the
        requirement that at  time $t_{I}$ the outermost $\Delta$
        scale sections must be contained in the $0$ scale section
        $R_{n_{I},0}$ of $R_{n_{I}}$ space. This requirement is
        needed if one requires that, at time $t_{I},$ $R_{n_{I}}$
        space is experimentally indistinguishable from the usual
        continuum $R$ space.

        The post inflationary Hubble expansion and redshift are
        accounted for by a continued increase of $n$ and a
        corresponding expansion factor of $e^{\epsilon \Delta n}.$
        If $\gamma$ is the constant increase rate of $n(t)$ where
        $n(t+\gamma^{-1})=1+n(t),$ then the Hubble constant
        $H=\epsilon\gamma.$ The present value of the Hubble
        constant $71\pm 7\; km/sec/mpc$ \cite{Freedman} gives a
        numerical relation between $\gamma$ and $\epsilon.$

        The discrete nature of $n$ means that a plot of the
        recession velocity against distance should be a step
        function with steps of size $\epsilon$ and frequency
        $\gamma.$ Examination of the literature data
        \cite{Freedman} shows no such step function within the
        accuracy of the data.  The accuracy is such that the data
        sets a lower limit on $\gamma$ of a unit increase in $n$
        every $30-60$ million years.

        Probably the most important point of this work is to show
        the suitability of $R_{n}$ space and time as a framework
        for inflationary cosmology.  No physical theory was
        provided to justify the choice of the dynamical model.
        Future work includes the need to provide such a theory or
        to tie this model in with other work. In addition there
        are other outstanding issues.  One is that this work was
        limited to sequences of even $2n$ length.  The treatment
        needs to be extended to cover sequences of an odd length.
        Also extension to numerical bases other than binary needs
        to be considered.

        On a more speculative note one needs to investigate if the
        seed locations that correspond to two and one dimensional
        singularities in $R_{n_{0}}$ space can serve as seeds for
        galactic black holes in $R_{n_{I},0}$ space.

       \section*{Acknowledgements}
       This work was supported by the U.S. Department of Energy,
       Office of Nuclear Physics, under Contract No. W-31-109-ENG-38.


\begin{thebibliography}{99}

       \bibitem{Wigner}
        E. Wigner, \textit{Commum. Pure and Applied Math.} {\bf 13} 001
            (1960), Reprinted in E. Wigner, {\it Symmetries and Reflections},
            (Indiana Univ. Press, Bloomington IN 1966), pp 222-237.

        \bibitem{BenLP}
       P. Benioff, \emph{Quantum Information Processing}, \textbf{1} 495, (2002), Archives:
       quant-ph/0210211; \emph{Foundations of Physics}, \textbf{32}, 989, (2002)

       \bibitem{BenTCTPMTEC}
       P. Benioff, Archives: quant-ph/0403209.

       \bibitem{Brandenberger}
       R. Brandenberger, Archives: hep-ph/9910410; hep-ph/0101119.

       \bibitem{Guthstein}
        A. H. Guth and P. Steinhardt, The Inflationary
        Universe, in \emph{The New Physics}, P. C. W. Davies,
        Editor, Cambridge University Press, New York, 1989,
        pp. 34-60.

        \bibitem{Watson}
        G. S. Watson, Archives: astro-ph/0005003.

        \bibitem{Freedman}
         W. L. Freedman, \emph{Phys. Rept.} \textbf{333} 13-31, (2000),
         Archives: Astro-ph/9909076.

         \bibitem{Guth}
         A. H. Guth, \emph{Phys. Rev. D}, \textbf{23}, 347-356,
         {1981}.

         \bibitem{Linde}
         A. D. Linde, \emph{Physics Letters}, \textbf{108B}
         389-393, (1982).

         \bibitem{Bahcall}
         N. Bahcall, J. P. Ostriker, S. Perlmutter, P. J.
         Steinhardt, \emph{Science}, \textbf{284} 1481-1488,
         (1999).







        \end{thebibliography}
        \end{document}